# SCALABILITY AND OPTIMISATION OF A COMMITTEE OF AGENTS USING GENETIC ALGORITHM

T. Marwala[*], P. De Wilde[*], L. Correia[**], P. Mariano[**], R. Ribeiro[**] V. Abramov[***], N. Szirbik[***], J. Goossenaerts[***]

[*]Electrical and Electronic Engineering Department, Imperial College, Exhibition Road, London SW7 2BT, England

[**]FCT- Dept. Informatics, Universidade Nova Lisboa, 2825-114 Monte Caparica, Portugal

[***]Information & Technology Department, Faculty of Technology Management, Eindhoven University of Technology, The Netherlands

**ABSTRACT**

A population of committees of agents that learn by using neural networks is implemented to simulate the stock market. Each committee of agents, which is regarded as a *player* in a game, is optimised by continually adapting the architecture of the agents using genetic algorithms. The committees of agents buy and sell stocks by following this procedure: (1) obtain the current price of stocks; (2) predict the future price of stocks; (3) and for a given price trade until all the players are mutually satisfied. The trading of stocks is conducted by following these rules: (1) if a player expects an increase in price then it tries to buy the stock; (2) else if it expects a drop in the price, it sells the stock; (3) and the order in which a player participates in the game is random. The proposed procedure is implemented to simulate trading of three stocks, namely, the Dow Jones, the Nasdaq and the S&P 500. A linear relationship between the number of players and agents versus the computational time to run the complete simulation is observed. It is also found that no player has a monopolistic advantage.

## 1. Introduction

In this work a population of committees of agents is used to simulate the stock market. Each committee of agents is viewed as a player in a game. These players compete and interact. The committee of agents [1] are optimised using genetic algorithm [2]. Perrone and Cooper [1] introduced a committee of networks, which optimises the decision-making of a population of networks. They achieved this by assuming that the trained networks were available and then assigning to each network a weighting factor, which indicates the contribution that the network has on the overall decision of a population of networks. The limitation of their proposal is that in a situation where the problem is changing such as the stock market the method is not flexible enough to allow the dynamic evolution of the population of networks. This paper seeks to fill the limitation on the committee procedure by ensuring that the individual networks that make the committee are allowed to dynamically evolve as the problem evolves. This is achieved by employing genetic algorithms [2]. The parameters defining the architecture of the networks, such as the number of hidden units, which form a committee, are defined as design variables and are allowed to evolve as the problem evolves. The network attributes that are fit to survive replace those that are not fit. On implementing genetic algorithm to select the fit individuals three steps are followed [2]: (1) crossover of network attributes within the population; (2) Mutation of each individual attributes; (3) and reproduction of the successful attributes. The simple crossover, the binary mutation and roulette wheel reproduction techniques are used. Finally the proposed technique is implemented to simulate the trading of three stocks. The scalability of the number of agents and players in the simulations with respect to computational time are investigated. The evolution of the complexity of the simulation as the players participate in more trading is also investigated.

## 2. Structure of the simulation technique

The structure that is proposed consists of committees of agents forming a player in the stock market. Each agent contributes equally to the overall decision-making of the player. The simulation framework consists of a population of these players that compete for fixed number of stocks. The agents learn using neural networks. The schematic illustration of the committee of agents is shown in Figure 1. The structure of each agent evolves using genetic algorithm such that its contribution to the overall function of a committee adapts to the evolutionary time-varying nature of the problem. The characteristics of the agents that evolve are the number of hidden units. The numbers of hidden units are constrained to fall within a given space, in this study 1 and 10. The output activation function is randomly chosen to be either linear or logistic [3].

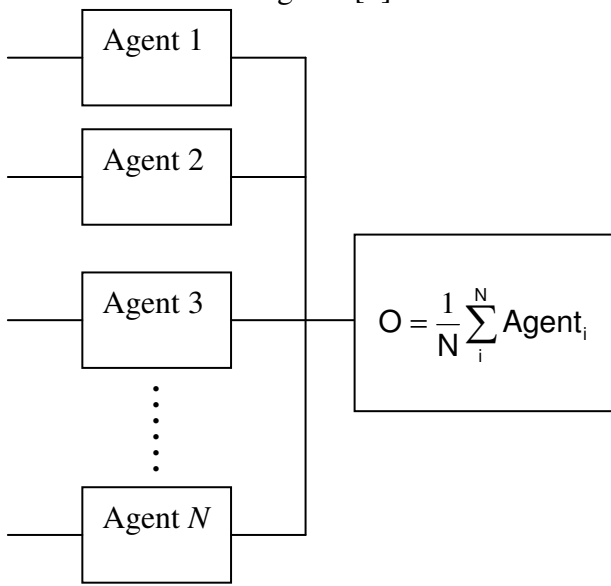

Figure 1: Schematic diagram representing the committee of agents.

Each committee of agents known as a player has trade stocks with other players. When prices of stocks are announced, the players trade by following these rules:

- Once a price is announced, the committees look at the current price and the future price of stocks. The future price is determined from the agents that that learn using neural networks. For a player, the predicted price is the average of the prediction of each agent within that particular player.
- If the predicted price of a stock is lower than the current price, then the player tries to sell the stock. If the predicted price for the stock is higher than the current price, then the committee tries to buy the stock.
- At any given stage the committee is only prepared to sell the maximum of 40% of the volume of stocks it has.
- The amount of stocks that a committee buys or sells depends on, amongst other factors, the predicted price. If the predicted price of a particular stock is x-% higher than the current price, the committee tries to acquire x-% of the volume available on the market of that particular stock.

This simulation is started by choosing the number of players that participate in the trading of stocks together with the number of agents that form a player. Then the agents are trained by randomly assigning the number of hidden units to fall in the interval [1 10]. The output activation units are randomly chosen to be linear or logistic. The agents are trained using the data from the previous 50 trading days. The trained agents are grouped into their respective players and are then used predict the next price given the current price. The players trade stocks using the rules that are be described later on in the paper. The schematic diagram of the simulation is shown in Figure 2.

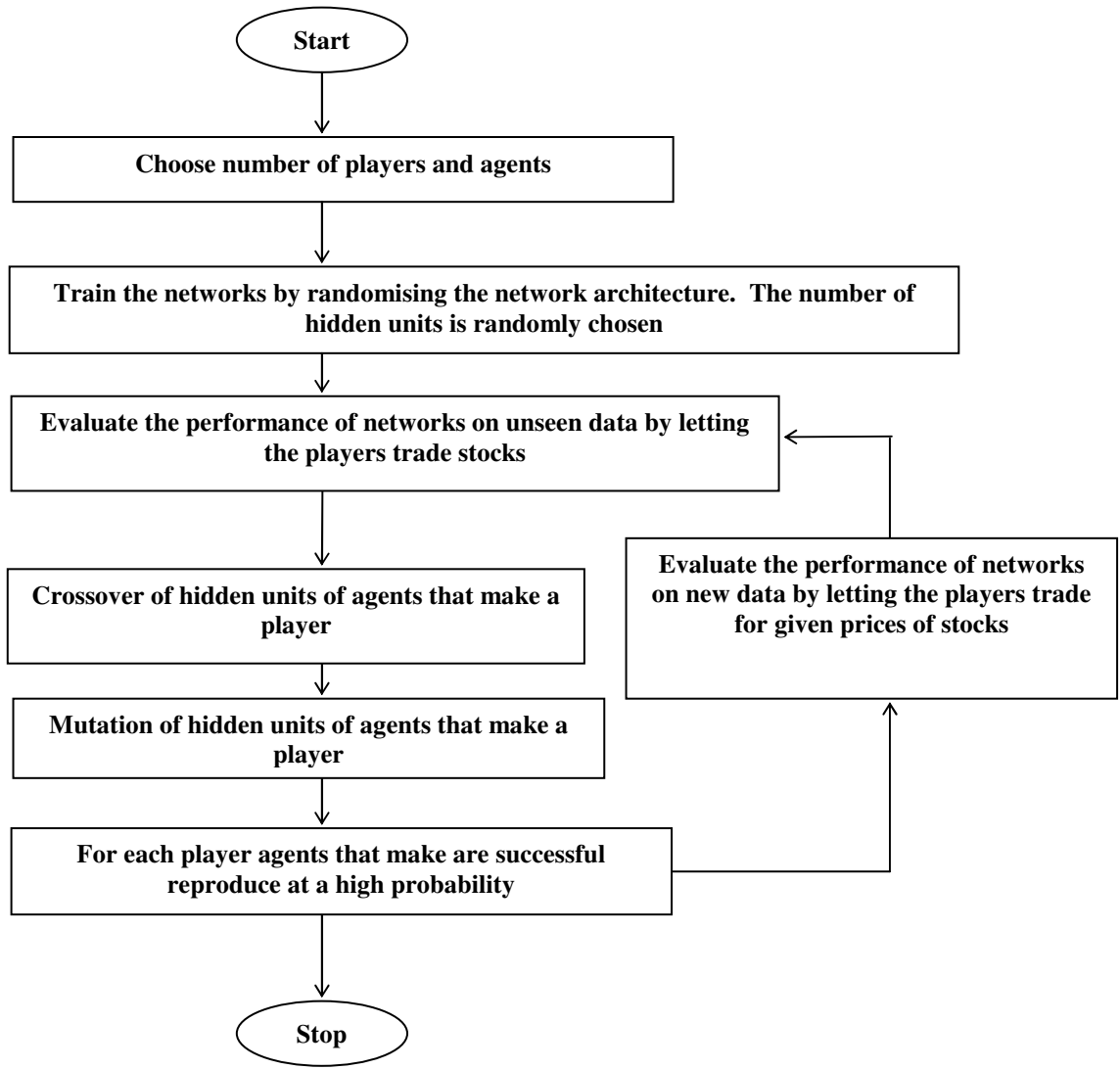

Figure 2: Schematic diagram of the trading simulation.

After 50 days have elapsed the performance of each agent is evaluated and the number of hidden units are transformed into 8 bits and within each player exchange bits with other players, a process called crossover. Thereafter, the agents mutate at low probability. The successful agents are duplicated while the less successful ones are eliminated. Then the networks are retrained again and the whole process is repeated. When a price is announced trading of stocks is conducted until the consensus is reached [4]. At this state, the overall wealth of the committees does not increase as a result of trading.

**3. Introduction to Genetic Algorithms**

Genetic algorithms were inspired by Darwin's theory of natural evolution. In natural evolution, members of the population compete with each other to survive and reproduce. Evolutionary successful individuals reproduce while weaker members die. As a result the genes that are successful are likely going to spread within the population. This natural optimization method is used in this paper to optimize the decision of a committee of agents in Figure 1. The basic genetic algorithm suggested by Holland [4] is implemented. The algorithm acts on a population of binary-string chromosomes. These chromosomes are obtained by using the Gray algorithm [5]. Each of these strings is discretised representation of a point in the search space. Here we are searching for the most optimum combination of architectures that form a committee and that give the least errors. Therefore the objective function is the error given by committee of agents. On generating a new population three operators are performed: (1) crossover; (2) mutation; (3) and reproduction. Just like in natural evolution, the probability of mutation occurring is lower than that of crossover or reproduction. The crossover operator mixes genetic information in the population by cutting pairs of chromosomes at random points along their length and exchanging over the cut sections. This operator has a potential of joining successful operators together. Simple crossover [4] is implemented in this paper. The mutation operator picks a binary digit of the chromosomes at random and inverts it. This has the potential of introducing to the population new information. Reproduction takes successful chromosomes and reproduces them in accordance to their fitness function. The fit parameters are allowed to reproduce and the weaker parameters are eliminated. This is done using the roulette wheel [2] procedure.

## 4. Application: Simulating the stock market

The example that is considered in this study is the trading of three stocks, namely: (1) the Dow Jones; (2) NASDAQ; (3) and S&P 500. The time-histories of the stocks are downloaded from the Internet and used to train agents. For a given set of price of these stocks the committee of agents predict the future prices of stocks. The following procedure is followed:

1. Using the time histories of the stocks generate N numbers of players trading M number of stocks. Each player consists of k-trained networks that form a committee.
2. Given the current price of $p_{mt}$ of the $m^{th}$ stock the at time t, the $n^{th}$ player predicts the price of the $m^{th}$ stock $p_{mn(t+1)}$ at time t+1.
3. Evaluate the change in prices $\Delta p_{mnt} = \frac{(p_{mn(t+1)} - p_{mnt})}{p_{mnt}}$.
4. For the $n^{th}$ player $m^{th}$ stock, calculate the decision factor $df_{mnt} = \Delta p_{mnt} Q_{mi}|_{i=1, i \neq m}$ where Q is the maximum quantity of stocks available for the $m^{th}$ stock.
5. Randomly choose the order in which the players are allowed to play.
6. For player m if $\|\max(df_{mnt})\| < \|\min(df_{mnt})\|$ then sell else buy the stock.
7. Repeat 3
8. Set t = t+1;
9. Go to 4.

It should be noted that on implementing this procedure the total number of stocks available is kept constant [6]. The sample results showing the net worth of players are shown in Figure 5. This figure indicates that sometimes the players with successful strategies do not necessarily dominate indefinitely. This is due to the fact that strategies that are successful in one time frame are not necessarily successful at a later time.

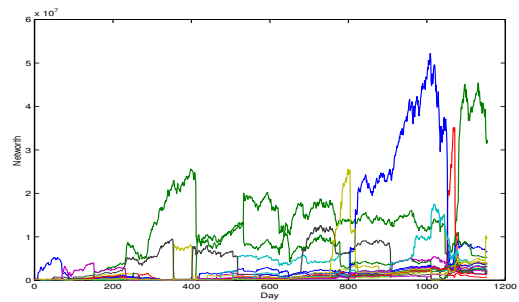

Figure 3: The net worth of each player versus trading days.

## 5. Scalability

In this section some aspect about the scalability of the simulations are studied. It is found that the method proposed is scalable. However, generally it was observed that the computational time increases with the increase in number of agents and players. A linear relationship exists between the average computational time taken to run the complete simulation and the number of players as well as the number of agents that form a player. This may be seen in Figures 4 and 5.

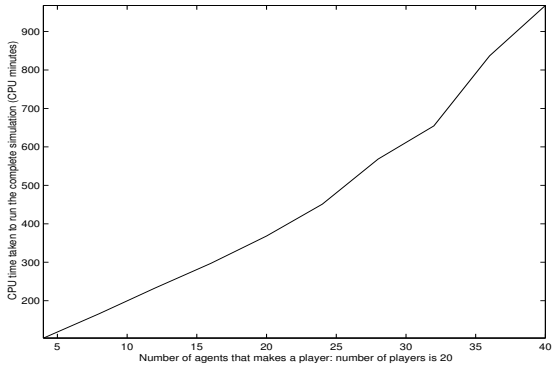

Figure 4: CPU time versus the number of players.

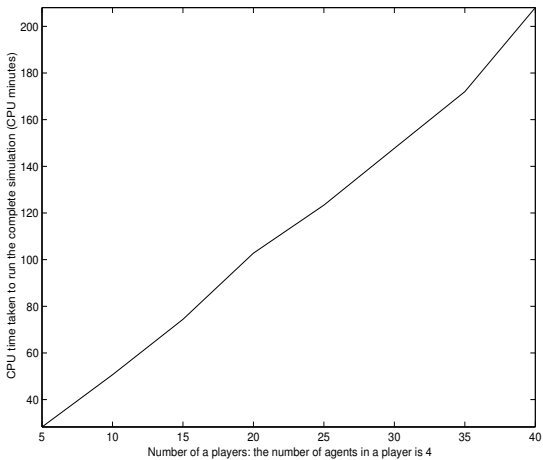

Figure 5: CPU time versus the number of agents.

## 6. Complexity

In this section we present a study of the complexity of the populations of agents that make players of the game. In this paper we view complexity as the measure of a degree of variation in a population of agents. Each species of agents form a dimension in space. Each dimension has a variation indicating the level of complexity of a population of that species. In mathematical for this definition of complexity may be written as:

$$Complexity = \sum_{i=1}^{N} \sigma_i e_i \qquad (1)$$

Figure 6 shows the average number of hidden units of an agent that forms a player versus the number of training conducted. This figure shows that as the system evolves the number of hidden units for a given player steadily decreases and stabilizes around 3.

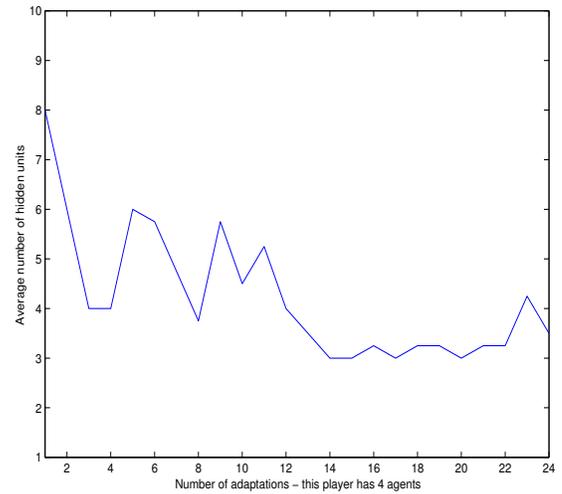

Figure 6: Average number of hidden units of an agent that forms a player versus the number of training conducted.

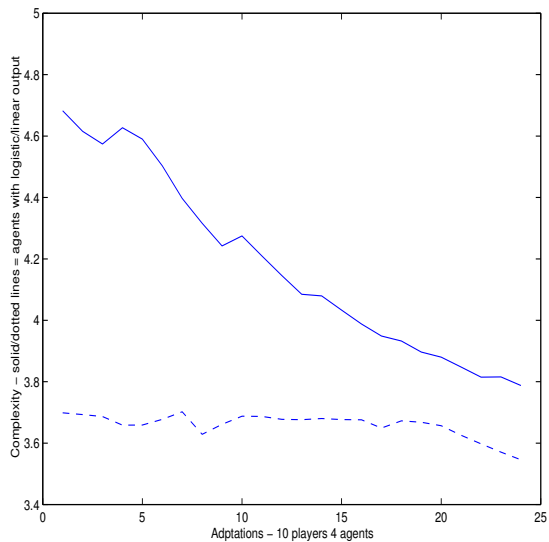

Figure 7: Complexity of species made of linear and logistic output functions versus the number of training conducted. Solid: Linear; Dashed: Logistic

The graph showing the complexity of two species i.e. logistic and linear output activation functions versus time of training is shown in Figure 7. This figure shows that on average the complexity of the agents with linear output units decreases as the time elapses. The complexity of the agents with logistic output units is approximately constant.

## 7. Conclusion

A simulation of the stock market was successfully conducted. It is found that the number of players and agents that form a player that participate in the trading game are directly proportional to the computational time taken to run the simulation. It is further observed that no player has the monopolistic advantage on the prediction of the stock market. The simulation also shows that as the time of the trading elapses, the complexity of the players decreases. This is due to the fact that as the time of trading elapses the players become more adapted to the time-varying nature of the problem, thereby developing common features. Optimising a committee of agents is found to be a feasible approach to modelling a player in the stock market.

## Acknowledgement

This work has been carried in the framework of research project No. IST-1999-10304 supported by Commission of the European Communities, European Union.